\documentclass[twocolumn,prd,showpacs,preprintnumbers,amsmath,amssymb,floatfix]{revtex4}

\usepackage{graphicx}

\usepackage{bm}
\usepackage{amsfonts}
\usepackage{lineno,hyperref}
\usepackage{array}
\usepackage{float}
\usepackage{epstopdf}
\usepackage{mathrsfs}
\usepackage{color}

\begin{document}

\title{\bf Generalized Ghost Dark Energy in DGP Model}

\author{Mahasweta Biswas}
\email{mahasweta.rs2014@math.iiests.ac.in} \affiliation{Department of Mathematics,
Indian Institute of Engineering Science and Technology, Shibpur,
Howrah 711103, West Bengal, India}

\author{Shounak Ghosh}
\email{shounak.rs2015@physics.iiests.ac.in} \affiliation{Department of Physics,
Indian Institute of Engineering Science and Technology, Shibpur,
Howrah 711103, West Bengal, India}

\author{Ujjal Debnath}
\email{ujjaldebnath@gmail.com} \affiliation{Department of Mathematics,
Indian Institute of Engineering Science and Technology, Shibpur,
Howrah 711103, West Bengal, India}

\begin{abstract}
In 2000, Giorgi Dvali, Gregory Gabadadze and Massimo Porrati (Dvali et al. 2000) was
proposed a new braneworld model named as DGP model, having two branches
with $(\epsilon=+1)$ and $(\epsilon=-1)$. Former one  $(~\epsilon=+1)$ known as the accelerating
branch, i.e. accelerating phase of the universe can be explained without adding
cosmological constant or Dark
energy, whereas later one represents the decelerating branch. Here we have
investigated the behavior of decelerating branch $(i.e.~\epsilon=-1)$ of DGP
model with Generalized Ghost Dark Energy (GGDE). Aim of our study to find a stable solution of the universe in DGP model. To find a stable solution we have studied the behavior of different cosmological parameters such as Hubble
parameter, equation of state (EoS) parameter and deceleration parameter with
respect to scale factor. Then we have analysed the $\omega_{D}-\acute{\omega_{D}}$ to confirm no freezing region of our present study
and point out thawing region. Furthermore we have checked the gradient of
stability by calculating the squared sound speed. Then we extend
our study to check the viability of this model under investigation
through the analysis of statefinder diagnosis parameters for the
present cosmological setup.
\end{abstract}
\date{}
\maketitle

\section{Introduction}

At the beginning of $20^{th}$ century Einstein constructed his famous general theory
of relativity which can be considered as the most successful theory to understand
the structure of our universe. Einstein himself believed that the universe
is stationary, but after the success of Hubble's law, expansion of the universe has
been confirmed and Einstein abandoned his concept of static universe from his theory.
Again in 1998 results of type Ia supernova explosion confirmed that our universe is
not only expanding but also in a phase of acceleration \cite{Riess1998,Riess1999,Perlmutter1999}.
This accelerated phase of the present universe is one of the most active topic of
research in cosmology. Though the reason behind this expansion with an
acceleration still a burning topic to the researchers. Researcher pointed out that
there must be a hidden source of energy which might be responsible for this phase.
This energy is termed as Dark Energy (DE) in literature. It has been assumed that
$73\%$ of the total energy of the universe is dark energy having vary high negative
pressure causing this acceleration. This DE model turned out to be the most promising
hypothesis to explain this current phase of acceleration. The simplest model of DE
is the Einstein's cosmological constant $\Lambda$. Though Einstein abandoned this
constant from his field equations to make his equations consistent with the Hubble's
law, but in order to explain the phase of  acceleration of the universe this constant
has to reappear in its same form but with different point of view.  This constant is
the key ingredient of the $\Lambda$-CDM model. In this model the equation of state
has the value $\omega_{DE}=-1$ ($\omega_{DE}$ is defined as the ratio between the
pressure and energy density of DE). In this scenario, the universe behaves asymptotically
with a de Sitter universe. Although the $\Lambda$-CDM model is consistent very well with
all observational data but it faces with the problems of fine tuning and coincidence \cite{Weinberg1989,Sahni2000,Carrall2001,Peebles2003,Padmanabhan2003,Tsujikawa2006}.
In order to rectify the problems plagued with $\Lambda$-CDM model various dark energy
models have been proposed time to time such as, chaplygin gas \cite{Kamenshchik2001,Bento2002,Zhang2006},
holographic \cite{Hsu2004,Li2004}, new agegraphic \cite{Cai2008}, polytropic gas \cite{Karami2009},
pilgrim \cite{Wei2012,Sharif2013,Sharif2013a}. In order to justify the source of accelerating
expansion (i.e. nature of DE) of the universe, two different approaches have been adopted.
One way to modify the geometry part of Einstein-Hilbert action (termed as modified theories of gravity)
for discussion of expansion phenomenon \cite{Sharif and Rani,Linder,Brans and Dicke,Dutta}
and other is propose to the different forms of DE called Dynamical DE models.

Each of the dark energy models have many hidden and unique features which always make a
challenging situation to the researchers. In general most of the dark energy models have
required an extra degree of freedom to explain this present phase of the universe \cite{Sheykhi2009,Sheykhi2009a,Sheykhi2010,Karami2011,Sheykhi2011c,Jamil2011}. This extra
term might creates inconsistency in the results. So for a desirable DE model one must resolve
the problem without adopting any new degree of freedom or any kind of extra parameter. In
order to do so a new model of DE known as Veneziano Ghost Dark Energy or simply Ghost Dark
Energy (GDE) has been proposed by  \cite{Urban2010,Urban2009,Urban2009a,Urban2009b,Ohta2011}.
This GDE attracts the attention of researcher as its energy density ($\rho_D$) depends linearly
on the Hubble parameter$(H)$ such as $\rho_D=\alpha H$,  where $\alpha$ is a constant.

With the consideration of this Veneziano ghost field, the $U(1)$ problem in low
energy compelling theory of Quantum Choromodynamics (QCD) has been resolved \cite{Witten1979,Veneziano1979,Rosenzweig1980,Nath1981,Kawarabayashi1980,K.Kawarabayashi1981,K.Kawarabayashi1981a}.
The $U(1)$ problem describes as the Lagrangian of QCD has in the massless limit, a
global Chiral $U(1)$ symmetry, which does not seems to be reflected in the
spacetimes of light pseudo scalar mesons. The ghost has no contribution to
 the vacuum energy density, which is proportional to $\Lambda^{3}_{QCD}H$, where
$\Lambda_{QCD}$ is QCD mass scale and $H$ is Hubble parameter. This small
vacuum energy density expect to play an important role in the evolution of universe
\cite{Rong-Gen Cai}. Considering GDE model above issues can be explained smoothly but
GDE model faces the problem of stability \cite{Ebrahimi2011b,CaiarXiv}, which clearly
indicating that the energy density does not depends on $H$ explicitly rather it depends
on the higher order terms of $H$ too, which is referred as Generalized Ghost Dark Energy
(GGDE) model. In GGDE model, the vacuum energy of the Ghost field can be taken as a
dynamical cosmological constant \cite{F.R.Urban1,F.R.Urban11,F.R.Urban111,F.R.Urban1111,Ohta}.
In the ref. \cite{A.R.Zhitnitsky}, the author discussed the contribution of the
Veneziano QCD ghost field to the vacuum energy is not exactly of order
$H$ and a sub-leading term $H^{2}$ appears due to the fact that the
vacuum expectations value of the energy momentum tensor is conserved
in isolation \cite{Maggiore2011}. Then the vacuum energy of the ghost field
can be written as $H+O(H^{2})$, where the sub-leading term $H^{2}$
play a crucial role in the early stage of universe evolution action as
the early DE \cite{Rong-Gen Cai}. The density of this generalized ghost
dark energy (GGDE) reads \cite{Rong-Gen Cai} as,
\begin{equation}\label{1}
\rho_{D}=\alpha H+\beta H^{2},
\end{equation}
where $\alpha$ and $\beta[energy]^{2}$ are the constant parameter of the model,
which should be determined.

The present work has been performed in the DGP braneworld model \cite{Gia Dvali}, which was proposed by Dvali, Gabadadze and Porrati in
2000. In this braneworld scenario our universe has been considered to be brane embedded
in higher dimensional spacetime. There are lot of works available in literature on
higher dimensional gravity especially in brane cosmology \cite{Antoniadis,Randal}.
This model indicates the existence of a $4+1$ dimensional Minkowski space, within which
ordinary $3+1$ dimensional Minkowski space is embedded. In DGP Gravity, the parameter
$\epsilon=\pm1$ correspond to two branches of DGP model. The solution with $\epsilon=+1$
leads to a self accelerating branch, for this branch dark energy is no longer be required
to describe the accelerated phase of the present universe  \cite{Koyama(2007)}. Whereas
$\epsilon=-1$ corresponds to the solution for normal branch, where it has
been claimed that dark energy is the only responsible of this accelerated
phase of expansion of the universe \cite{Lazkozet al.2006}. To overcome this problem different investigation have been attempted
to discuss DE model in DGP theory, a cosmoligical constant \cite{Lue and starkman 2004,Sahni and Shtanov 2003,Lazkoz}, a quintessence perfect fluid \cite{Chimento}, a scalar field \cite{Zhang}
or chaplygin gas \cite{Bouhmadi} and HDE \cite{Wu,Liu}. Motivating from these
works for different DE models we have studied of GGDE model under DGP gravity and we
are able to obtain physically acceptable and stable results in favor of the current
phase of the universe.

The Universe's expansion rate can be explained by the Hubble
parameter $H =\frac{\dot{a(t)}}{a(t)}$, where $a(t)$ is the cosmic scale factor of the
Universe and the over dot on $a$ stands for the time derivative of it.
On the other hand, the deceleration parameter $(q)$ also describes the rate of the deceleration or acceleration
of the Universe,
\begin{equation}\label{2}
q=-\frac{a\ddot{a}}{\dot{a}^{2}}=-\frac{\ddot{a}}{a H^{2}}.
\end{equation}

The hubble parameter ($H$) and deceleration parameter ($q$) are well known cosmological
parameters which explain the evolution of the universe. However these two parameters cannot discriminate
among various DE models. In this context, Sahni et al. \cite{Sahni2003} and Alam et al. \cite{Alam2003} have introduced
a new geometrical diagnostic pair $(r,s)$ known as statefinder parameter. Which can be
derived using the scale factor $(a)$ and its time derivatives upto third order. The statefinder
parameter is completely geometric in nature as it deduced directly from the spacetime metric.
These parameters are more dependable than any other parameters to study
the physical acceptability of any DE models and to distinguish between them. So the statefinder parameters can be written
as,
\begin{equation}
 r=\frac{\dddot{a}}{a H^{3}}~~ {\text{and}}~~
s=\frac{r-1}{3(q-\frac{1}{2})}. \label{3}
\end{equation}
For a flat $\Lambda CDM$ model the pair has a fixed value $\{r,s\}=\{0,1\}$.

So our present study in organised in the following manner. In Sec. \ref{sec2} we have discussed basic mathematics of DGP braneworld model.
The behavior of different cosmological parameters such as Hubble parameter, deceleration parameter and EoS parameter has been describe in
Sec. \ref{sec3}. Then we have studied $\omega_{D}-\acute{\omega_{D}}$ analysis in Sec. \ref{sec4}. In Sec. \ref{sec5} we check the stability of our model. Furthermore we have described the statefinder diagnosis in Sec. \ref{sec6}. Finally we have concluded some of the important results in Sec. \ref{sec7}.

\section{DGP Braneworld Model}\label{sec2}
In this section we have extended our study under the braneworld scenario,
the five dimensional spacetime of our universe can be realized as a 3-brane
embedded spacetime. Dvali- Gabadadze-Porrati (DGP) proposed a new version of
this braneworld scenario in which our four dimensional Friedman-Robertson-Walker
universe embedded in five dimensional Minkowski spacetime. The usual gravitational
laws in this scenario can be obtained by the addition of the action with the
Einstein-Hilbert action term estimated with the brane inherent curvature.
Existence of this specific term in the action is induced due to quantum
corrections arising from the bulk gravity and its coupling with matter
living on the brane. In this model the cosmological evolution on the brane
has been described by an effective Friedman equation which includes the
non-trivial bulk effects onto the brane.

So the modified Firedman equation for an isotropic and homogenous universe
related to our model can be written as follows \cite{Koyama 2008,Li et al 2004,Li et al 2011}

\begin{equation}\label{4}
H^{2}-\frac{\epsilon}{r_{c}}\sqrt{H^{2}+\frac{k}{a^{2}}}=\gamma \rho-\frac{k}{a^{2}},
\end{equation}
where $\gamma=\frac{8\pi G}{3}$. The total cosmic
fluid energy density $\rho$ can be written as $\rho=\rho_{m}+\rho_{D}$, where $\rho_{D}$
is the energy density of DE and $\rho_{m}$ is that for DM on the brane, $k$
represents the curvature parameter, $k$ can have the values $k=-1,0,1$
corresponds to open, flat and closed universe respectively in maximally
symmetric space on the brane.

 Here $r_{c}=\frac{M_{pl}^{2}}{2 M_{5}^{3}}=\frac{G_{5}}{2 G_{4}}$
denotes the cross over scale length which can be defined as the upper
limit of the length at which the universe begins to dominate by higher
dimensions in late time where the Hubble parameter leads towards $H\approx\frac{1}{r_c}$.

In flat DGP braneworld $(k=0)$, the Friedmann eqauion of Eq. \ref{1} reduces to
\begin{equation}\label{5}
H^{2}-\frac{\epsilon}{r_{c}}H =\gamma (\rho_{m}+\rho_{D}).
\end{equation}

Now we define the
dimensionless density parameters as usual
\begin{equation}\label{6}
\Omega_{m}=\frac{\rho_{m}}{\rho_{cr}} ~\text{and}~ \Omega_{D}=\frac{\rho_{D}}
{\rho_{cr}}
\end{equation}
where $\rho_{cr}=\frac{H^{2}}{\gamma}$, which is the critical
energy density. Thus the Friedmann equation reduces to the form

\begin{equation}\label{7}
1-\epsilon \sqrt{\Omega_{r_{c}}}=\Omega_{m}+\Omega_{D},
\end{equation}
where we define-
\begin{equation}\label{8}
\Omega_{r_{c}}=\frac{1}{H^{2} r_{c}^{2}}.
\end{equation}
In this present study we have used the non interacting conservation equation for effective DE and DM can be
reduced to the following forms-
\begin{equation}\label{9}
\dot{\rho}_{D}+3H \rho_{D}\left(1+ \omega_{D}\right)=0,
\end{equation}
\begin{equation}\label{10}
\dot{\rho}_{m}+3 H \rho_{m}=0\Rightarrow \rho_{m}=\rho_{m0}a^{-3},
\end{equation}
where $\omega_{D}=\frac{p_{D}}{\rho_{D}}$ is the equation of state parameter of DE.

Now substituting the values of $\rho_{D}$ and $\rho_{m}$ from the Eqs.
(\ref{1}) and (\ref{10}) in the Friedmann Eq. (\ref{5}), we get

\begin{equation}\label{11}
(1-\beta\gamma)H^{2}-(\frac{\epsilon}{r_{c}}+\alpha\gamma)H=\gamma \rho_{m0}a^{-3}.
\end{equation}

From the above equation, we have obtained the Hubble parameter $H(a)$ as
\begin{equation}\label{12}
H_{\pm}=\frac{(\frac{\epsilon}{r_{c}}+\gamma \alpha) \pm \sqrt{
(\frac{\epsilon}{r_{c}}+\gamma \alpha)^{2}+4\gamma (1-\gamma \beta) \rho_{m0}a^{-3}}}{2(1-\gamma \beta)}.
\end{equation}

We get two values of Hubble parameter H. The expansion of
Universe is denoted by $H_{+}$. Based on the observational results ignoring the
later one we are continue with $H_{+}$. We use H instead of $H_{+}$ throughout our
paper for simplicity.

\begin{equation}\label{13}
H_{+}=H=\frac{(\frac{\epsilon}{r_{c}}+\gamma \alpha) + \sqrt{
(\frac{\epsilon}{r_{c}}+\gamma \alpha)^{2}+4\gamma (1-\gamma \beta) \rho_{m0}a^{-3}}}{2(1-\gamma \beta)}.
\end{equation}

Now we define characterised scale factor
$a_{\star}$ to continue our present discussion  as

\begin{equation}\label{14}
a_{\star} \equiv \left(\frac{4 \gamma \rho_{DM0}(1-\gamma \beta)}{(\frac{\epsilon}{r_{c}}+\gamma \alpha)^{2}}\right)^\frac{1}{3}
=\left(\frac{4 \Omega_{DM0}(1-\gamma \beta)}{(\epsilon \sqrt{\Omega_{r_{c0}}}+\Omega_{D0}-\gamma\beta)^{2}}\right)^\frac{1}{3}.
\end{equation}

The transition point of the universe from the dust phase to present de
Sitter phase was represents by this characteristics scale factor. Following recent observational data of Planck collaboration \cite{Planck2016}, we take
the values of $\Omega_{D0}$, $\Omega_{m0}$ ,$\Omega_{r_{c0}}$ and $\beta$
to be $0.689$, $0.35$, $0.03$ and $-0.1$
respectively then we get $a_\star \sim 1$, this indicates that
the transition takes place just at present.

\subsection{Energy Density}
Now we obtain the energy density of dark energy by using Eq. (\ref{13}) as
\begin{eqnarray}\label{15}
&&\frac{2(1-\gamma \beta)}{(\frac{\epsilon}{r_{c}}+\gamma \alpha)}  \rho_{D}=\left(1+\sqrt{1+(\frac{a_{\star}}{a})^{3}}\right) \nonumber \\
&&\left[\alpha+\frac{\beta(\frac{\epsilon}{r_{c}}+\gamma \alpha)}{2(1-\gamma \beta)}\left(1+\sqrt{1+(\frac{a_{\star}}{a})^{3}}\right)\right].
\end{eqnarray}

\begin{figure}[thbp]
\centering
\includegraphics[width=0.35\textwidth]{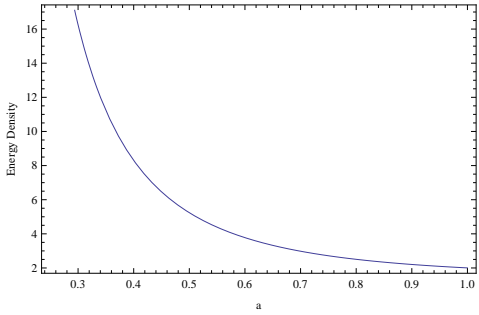}
\caption{Variation of energy density ($\frac{2(1-\gamma \beta)\rho_{D}}{(\frac{\epsilon}{r_{c}}+\gamma \alpha)}$) with $a$. }\label{fig1}
\end{figure}

Variation of $\rho_{D}$ with respect to scale factor $a$ has been shown in Fig.~\ref{fig1}.
From the figure it is clear that the energy density is decreasing with
the expansion of the universe.

\subsection{Cosmic time evolution}

\begin{figure*}[thbp]
\centering
\includegraphics[width=0.35\textwidth]{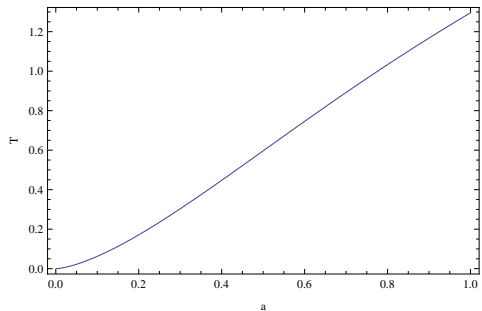}
\caption{Variation of cosmic time evolution $T (=\frac{(\frac{\epsilon}{r_{c}}+\gamma \alpha)(t-t_{i})}{2(1-\gamma \beta)})$  with $a$. }\label{fig2}
\end{figure*}

Now to describe the time evolution of the Universe,
we solve the Eq. (\ref{13}) analytically and we get
\begin{eqnarray}\label{16}
\frac{(\frac{\epsilon}{r_{c}}+\gamma \alpha)}{2(1-\gamma \beta)}(t-t_{i})
=-y^{3}+y^{3}\sqrt{1+y^{-3}}+\frac{3}{2} \ln y \nonumber \\
+\ln(1+\sqrt{1+y^{-3}}),
\end{eqnarray}
where $y=\frac{a}{a_\star}$ and $t_i$ represent the initial time when $a(t_i)=0$.
Now there are two different conditions for cosmic time evolution of the universe.
Firstly at early time of the universe $y\ll 1$ , the RHS of the Eq. (\ref{16})
indicates that $\frac{(\frac{\epsilon}{r_{c}}+\gamma \alpha)}{2(1-\gamma \beta)}
(t-t_{i}) \approx\ 2 y^\frac{3}{2}$, and for late time $y\gg 1$, the Eq. (\ref{16})
indicates that $\frac{(\frac{\epsilon}{r_{c}}+\gamma \alpha)}{2(1-\gamma \beta)}
(t-t_{i}) \approx\frac{3}{2} \ln y$. Variation of the cosmic time evolution with
respect to scale factor `$a$' is shown
in Fig.~\ref{fig2}. From the figure it is clear that the cosmic time evolution
is represent the present accelerated phase from early decelerating phase of the universe.

\section{Cosmological Parameters}\label{sec3}

The study of cosmological parameters is an important tool to describe
various properties of the universe. To describe these properties cosmological
parameters includes the parameterizations of some functions, as well as some
simple numbers. Actually these parameters are describing the global
dynamics of the universe, i.e. the expansion rate and the curvature. Nowadays
these parameters has been also studied with a great interest to describe the formation of
the universe from baryons, photons, neutrinos, dark matter and dark energy.
For any acceptable physical model these parameters play an important role.
We have studied some of the basic parameters such as Hubble parameter, EoS
parameter and deceleration parameter of our present GGDE model under DGP braneworld gravity.

\subsection{Hubble Parameter}

From the Eqs. (\ref{13}) and (\ref{14}) we observe that at early
time of the universe i.e at $a\ll a_\star$, $H\propto a^{-\frac{3}{2}}$,
which indicates the dust phase of the universe. For $a \gg a_\star$, i.e
at late time of the universe, $H=$ constant, which denotes the entry at
the de sitter phase at the later epoch. These values are perfectly right
which mentioned earlier that $a_\star$ represents the transition point
between the two epochs.

Variation of Hubble parameter of Eq. (\ref{13}) with respect cosmic scale
factor $(a)$ has been shown in Fig.~\ref{fig1}. The figure clearly indicating that
value of Hubble parameter is decreases with the evolution of the Universe.

\begin{figure}[thbp]
\centering
\includegraphics[width=0.35\textwidth]{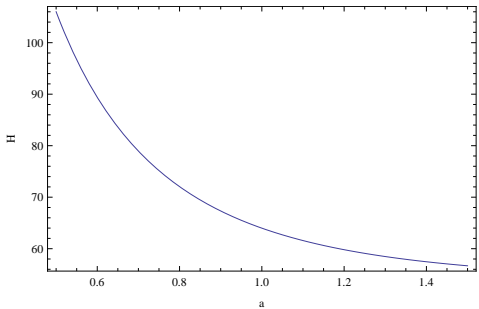}
\caption{Variation of $H$ with $a$ }\label{fig3}
\end{figure}

\subsection{EoS Parameter}

Now we check the behavior of equation of state $(\omega_{D})$ parameter in this model.
To evaluate the value of $(\omega_{D})$ we first take the time derivative of  Eq. (\ref{13})
after using Eq. (\ref{14}), we arrive as
\begin{eqnarray}\label{17}
\frac{\dot{H}}{H^{2}}=-\frac{3}{2}\frac{(\frac{a_{\star}}{a})^{3}}
{\sqrt{1+(\frac{a_{\star}}{a})^{3}} [1+\sqrt{1+(\frac{a_{\star}}{a})^{3}}]}.
\end{eqnarray}

From Eq. (\ref{9}) we get
\begin{eqnarray}\label{18}
\omega_{D}=-\frac{\dot{\rho_{D}}}{3H \rho_{D}}-1.
\end{eqnarray}

While taking time derivative of $\rho_{D}=\alpha H+\beta H^{2}$, we find
\begin{eqnarray}\label{19}
\frac{\dot{\rho_{D}}}{\rho_{D}}=\frac{\alpha+2 \beta H}{\alpha+\beta H}\frac{\dot{H}}{H}.
\end{eqnarray}

Hence the EoS $(\omega_{D})$ parameter for generalized ghost dark energy obtained as:
\begin{eqnarray}\label{20}
&&\omega_{D}=\left(\frac{a_{\star}}{a}\right)^{3}\left(\frac{1}{\sqrt{1+
(\frac{a_{\star}}{a})^{3}}}-\frac{1}{1+\sqrt{1+(\frac{a_{\star}}{a})^{3}}}\right) \nonumber \\
&&\left(\frac{\Omega_{D}+\beta\gamma[(\epsilon \sqrt{\Omega_{r_{c}}}-1)+X]}
{2\Omega_{D}+\beta\gamma[(\epsilon \sqrt{\Omega_{r_{c}}}-\Omega_{D}+\beta\gamma-2)+X]}\right)-1\nonumber\\
&=&
    \begin{cases}
      0, & \ a\ll a_{\star} \\
      -1, & a \gg a_{\star},
    \end{cases}
\end{eqnarray}
where $X=(\Omega_{D}-\beta\gamma+\epsilon \sqrt{\Omega_{r_{c}}})\sqrt{1+(\frac{a_{\star}}{a})^{3}}$.

\begin{figure}[thbp]
\centering
\includegraphics[width=0.35\textwidth]{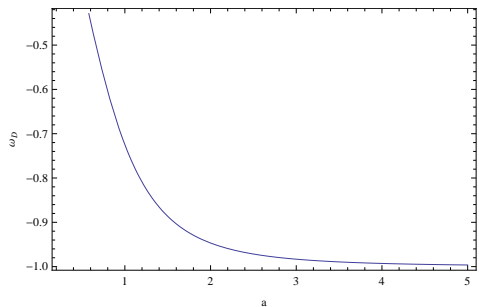}
\caption{Variation of EoS parameter with $a$. }\label{fig4}
\end{figure}

From this relation we can conclude that at late time the DE acts
like a cosmology constant due to asymptotic behavior of
$\omega_{D}$. Now we plot the figure of $\omega_{D}$ vs $a$ in
Fig.~\ref{fig4}. From the graph, we see that $\omega_{D}$ can never cross
$-1$, which is similar to quintessence behavior. EoS parameter
varies from zero at early time to $-1$ at late time.

Now taking time derivative of $\Omega_{D}$ we can obtain the equation
of motion for dimensionless GGDE density as-
\begin{eqnarray}\label{21}
\dot{\Omega_{D}}=-(\Omega_{D}-\beta\gamma)(\frac{\dot{H}}{H}).
\end{eqnarray}

Using the relation  $\dot{\Omega}_{D}=H\frac{d \Omega_{D}}{d \ln a}$ with Eq. (\ref{17}), we obtain-
\begin{eqnarray}\label{22}
\frac{d \Omega_{D}}{d \ln
a}=-\frac{3(\Omega_{D}-\beta\gamma)(\frac{a_{\star}}{a})^{3}}
{2\sqrt{1+(\frac{a_{\star}}{a})^{3}} [1+\sqrt{1+(\frac{a_{\star}}{a})^{3}}]}.
\end{eqnarray}

\begin{figure}[thbp]
\centering
\includegraphics[width=0.35\textwidth]{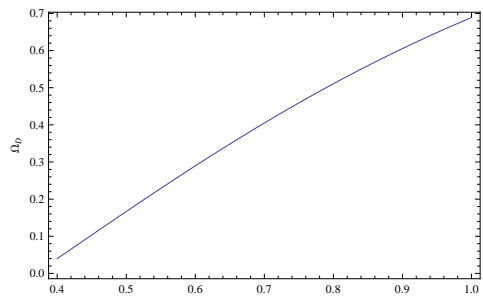}
\caption{The evolution of $\Omega_{D}$ versus with $a$.}\label{fig5}
\end{figure}

The extension of the dimensionless GGDE density $\Omega_{D}$ in terms of scale factor $a$ has
been shown in Fig.~\ref{fig5}. From the figure we observe that at early time
$\Omega_{D}\rightarrow 0$ and at late time $\Omega_{D}\rightarrow 1$, i.e. dark energy dominated.

We can also evaluate the equation of motion for $\Omega_{r_{c}}$. In order to do so first we
take time derivative of  Eq.~(\ref{8}), and then using Eq.~(\ref{17}) we get
\begin{eqnarray}
\frac{d \Omega_{r_{c}}}{d \ln
a}=-\frac{3 \Omega_{r_{c}}}{2}\frac{(\frac{a_{\star}}{a})^{3}}
{\sqrt{1+(\frac{a_{\star}}{a})^{3}} [1+\sqrt{1+(\frac{a_{\star}}{a})^{3}}]}. \label{23}
\end{eqnarray}

This is the equation of motion governing the evolution of GGDE under the framework of
DGP Gravity.

\subsection{Deceleration Parameter}
Deceleration parameter is one of the most important parameter which
measure expansion history of the universe. We already solved the value
of $\frac{\dot{H}}{H^{2}}$ in Eq. (\ref{17}), as a function of scale factor $a$
putting this value in Eq. (\ref{17}), we can easily estimate the value of $q$ in terms of $a$ as
\begin{eqnarray} \label{24}
q&=&-(1+\frac{\dot{H}}{H^{2}})\nonumber\\&=&-1+\frac{3}{2}\left(\frac{a_{\star}}{a}\right)^{3}\left[\frac{1}
{\sqrt{1+(\frac{a_{\star}}{a})^{3}}}-\frac{1}{1+\sqrt{1+(\frac{a_{\star}}{a})^{3}}}\right] \nonumber\\
&=&
    \begin{cases}
      \frac{1}{2}, & \ a\ll a_{\star} \\
      -1, & a \gg a_{\star}.
    \end{cases}
\end{eqnarray}

\begin{figure}[thbp]
\centering
\includegraphics[width=0.35\textwidth]{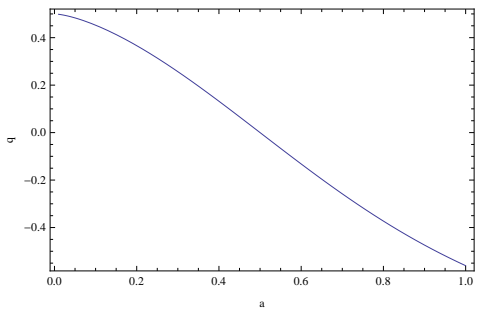}
\caption{Variation of Deceleration parameter with $a$. }\label{fig6}
\end{figure}

Deceleration parameter $q$ decreases monotonically from $\frac{1}{2}$ to −1,
which means that the expansion of the universe undergo a transition from
deceleration at early epoch to acceleration at present time.
Now we have plotted $q$ against scale factor for this model in Fig.~\ref{fig6}.
From the figure we analyze the behavior of deceleration parameter
corresponding to same fixed values of constrains. We see that for
increases of scale factor deceleration parameter was decreasing
and decreases to more negative values. Thus the negative value of
deceleration parameter demonstrates acceleration expansion of the
universe, which was totally perfect for GGDE phenomenon.

\section{$\omega_{D}-\acute{\omega_{D}}$ Analysis}\label{sec4}

To discuss the behavior of quintessence DE model, the $\omega_{D}-\acute{\omega_{D}}$
analysis was firstly proposed by Caldwell and Linder\cite{Caldwell2005}, where prime
denotes derivative with respect to $\ln a$. They also examined the limit of quintessence
model as showing thawing ($\acute{\omega_{D}}>0$  for $\omega_{D}<0$) and freezing
($\acute{\omega_{D}}<0$ for $\omega_{D}<0$) region by constructing $\omega_{D}-\acute{\omega_{D}}$
plane. The universe's expansion in freezing region is more accelerated as compared to thawing region.

To discuss the $\omega_{D}-\acute{\omega_{D}}$ analysis for this model, we compute
$\acute{\omega_{D}}$ from the Eq. (\ref{20}), by taking derivative with respect to $\ln a$.
Then the value of $\acute{\omega_{D}}$ is given below-

\begin{eqnarray}\label{25}
\acute{\omega_{D}}=\frac{1}{3}[(AB)^{2}-\frac{1}{C}AB-\frac{2\beta C}{\alpha+\beta C}B^{2}],
\end{eqnarray}
where
$A=\frac{2[\alpha+\beta[\frac{\epsilon}{r_{c}}+(\alpha\gamma+\frac{\epsilon}{r_{c}})\sqrt{1+(\frac{a_{\star}}{a})^{3}}]]}
{2\alpha+\beta[\frac{\epsilon}{r_{c}}+(\alpha\gamma+\frac{\epsilon}{r_{c}})\sqrt{1+(\frac{a_{\star}}{a})^{3}}-\alpha\gamma]}$,\\

$B=-\frac{3}{2}\frac{(\frac{a_{\star}}{a})^{3}}
{\sqrt{1+(\frac{a_{\star}}{a})^{3}} [1+\sqrt{1+(\frac{a_{\star}}{a})^{3}}]}$,\\

$C=\frac{(\frac{\epsilon}{r_{c}}+\gamma \alpha)}{2(1-\gamma \beta)}\left(1+\sqrt{1+(\frac{a_{\star}}{a})^{3}}\right)$.\\

\begin{figure}[thbp]
\centering
\includegraphics[width=0.35\textwidth]{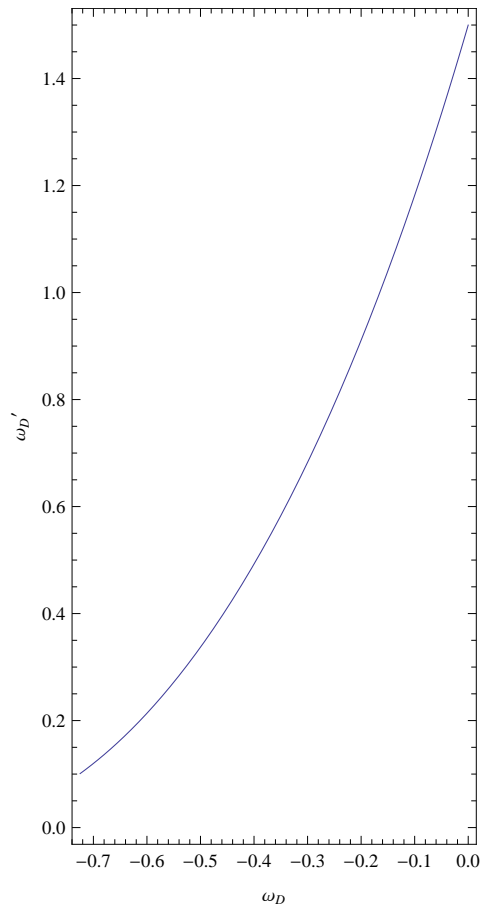}
\caption{Variation of $\acute{\omega_{D}}$ with $\omega_{D}$ }\label{fig7}
\end{figure}

Using this value of $\acute{\omega_{D}}$ of Eq.~(\ref{25}) and the value of $\omega_{D}$ from  Eq.~(\ref{20}), we
plot a graph between $\omega_{D}$ and $\acute{\omega_{D}}$ given
in Fig.~\ref{fig7}. From the figure, we observe that $\acute{\omega_{D}}$
decreases as $\omega_{D}$ decreases.

\section{Stability Analysis}\label{sec5}

In order to analyze the stability of GGDE model in this scenario,
we extract the square speed of sound which is given by
\begin{equation}\label{26}
v_{s}^{2}=\frac{dp}{ds}=\frac{\dot{P}}{\dot{\rho}}=\frac{\rho}
{\dot{\rho}}\dot{\omega_{D}}+\omega_{D}.
\end{equation}

Now putting the values of the parameters in right hand side of the above
equation we get the value of the sound speed as
\begin{eqnarray}\label{27}
v_{s}^{2}=\frac{(\frac{a_{\star}}{a})^{3}\{\beta\gamma(1-\epsilon \sqrt{\Omega_{r_{c}}})-\Omega_{D}\}}
{2\{1+(\frac{a_{\star}}{a})^{3}\}[\{\Omega_{D}-\beta\gamma(1-\epsilon \sqrt{\Omega_{r_{c}}})\}+\beta\gamma X]}.
\end{eqnarray}

\begin{figure}[thbp]
\centering
\includegraphics[width=0.35\textwidth]{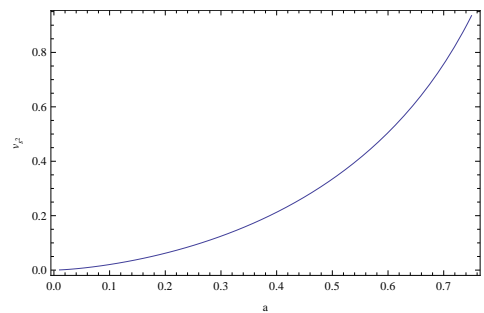}
\caption{Variation of speed of sound with $a$.}\label{fig8}
\end{figure}

We have shown the variation of sound speed $(v_s^2)$ with respect
to the scale factor $a$ in Fig.~\ref{fig8}. From the figure we have observed
that for GGDE model the sound speed remains positive and less than
$1$ throughout cosmic evolution.
Which suggest the stability of our model. Cai et al.
have claimed \cite{Cai2012} that for the best fitting results sub-leading
term must be negative. Following their argument we have also found
that for negative value of $\beta$ the model shows stability.

\section{Statefinder Diagnosis}\label{sec6}
To elaborate the phenomenon of GGDE in the accelerated expansion
of the universe we proposed many different type of GGDE model. It
is very important process to differentiate these model because of
that one can decide which one provides better explanation for the
current status of the universe. Since these parameters are
essential, Sahni \cite{Sahni2003} introduced two new dimensionless
parameters by combining the Hubble and deceleration parameter
which are called statefinder parameters, are one of the most
useful in geometrical tool in the sense that we can find the
distance of a given GGDE model from $\Lambda CDM$ limit. Now using
Eq.~(\ref{17}) and Eq.~(\ref{24}) in Eq.~(\ref{3}) eventually we obtain the pair of statfinder parameters as

\begin{eqnarray}\label{28}
&&r=1+3\frac{\dot{H}}{H^{2}}+\frac{\ddot{H}}{H^{3}}\nonumber\\
&=&-\frac{\{(\frac{a_{\star}}{a})^{6}-16 (\frac{a_{\star}}{a})^{3}-8\}
-4\{1+(\frac{a_{\star}}{a})^{3}\}^{\frac{3}{2}}\{2+(\frac{a_{\star}}{a})^{3}\}}
{4\{1+(\frac{a_{\star}}{a})^{3}\}^{\frac{3}{2}}\{1+\sqrt{1+(\frac{a_{\star}}{a})^{3}}\}^{2}},
\end{eqnarray}

\begin{eqnarray}\label{29}
s=\frac{1}{2}\frac{(\frac{a_{\star}}{a})^{6}}{(1+\sqrt{1+(\frac{a_{\star}}{a})^{3}})^{2}[1+(\frac{a_{\star}}{a})^{3}]}.
\end{eqnarray}

\begin{figure}[thbp]
\centering
\includegraphics[width=0.35\textwidth]{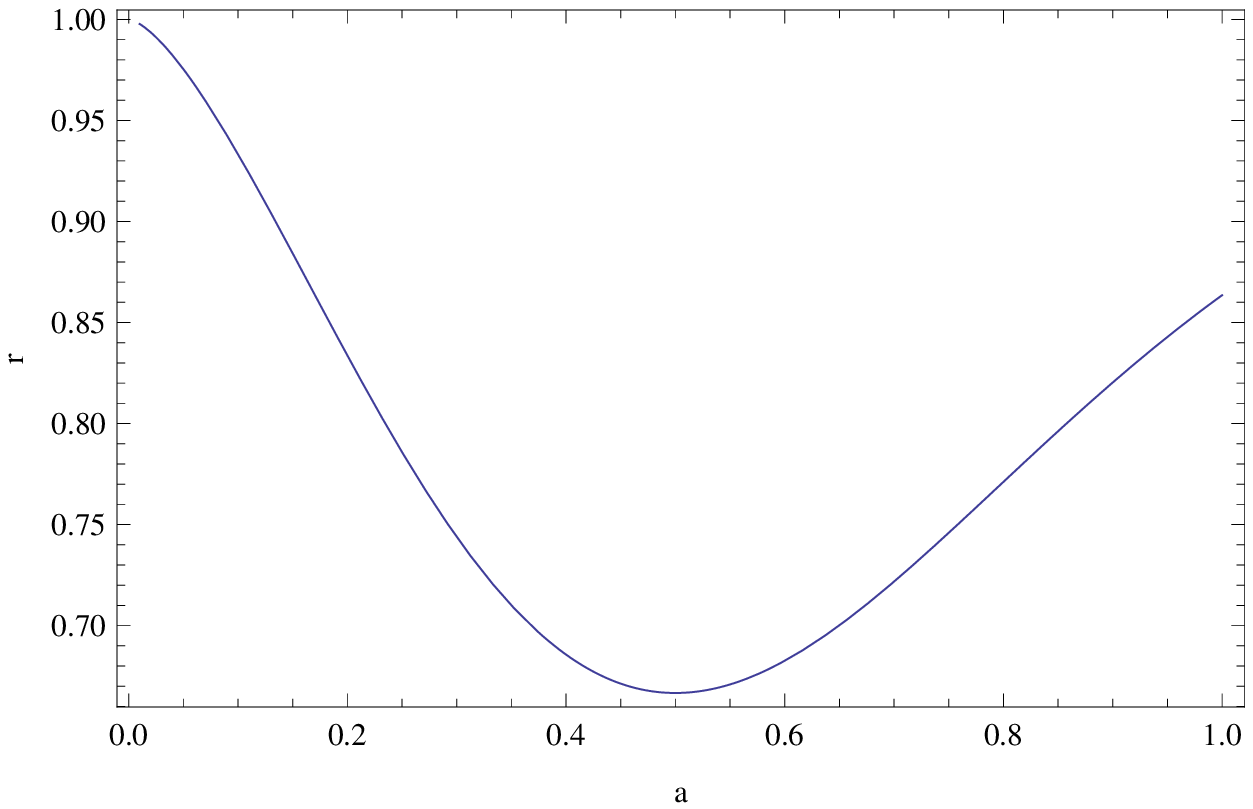}
\includegraphics[width=0.35\textwidth]{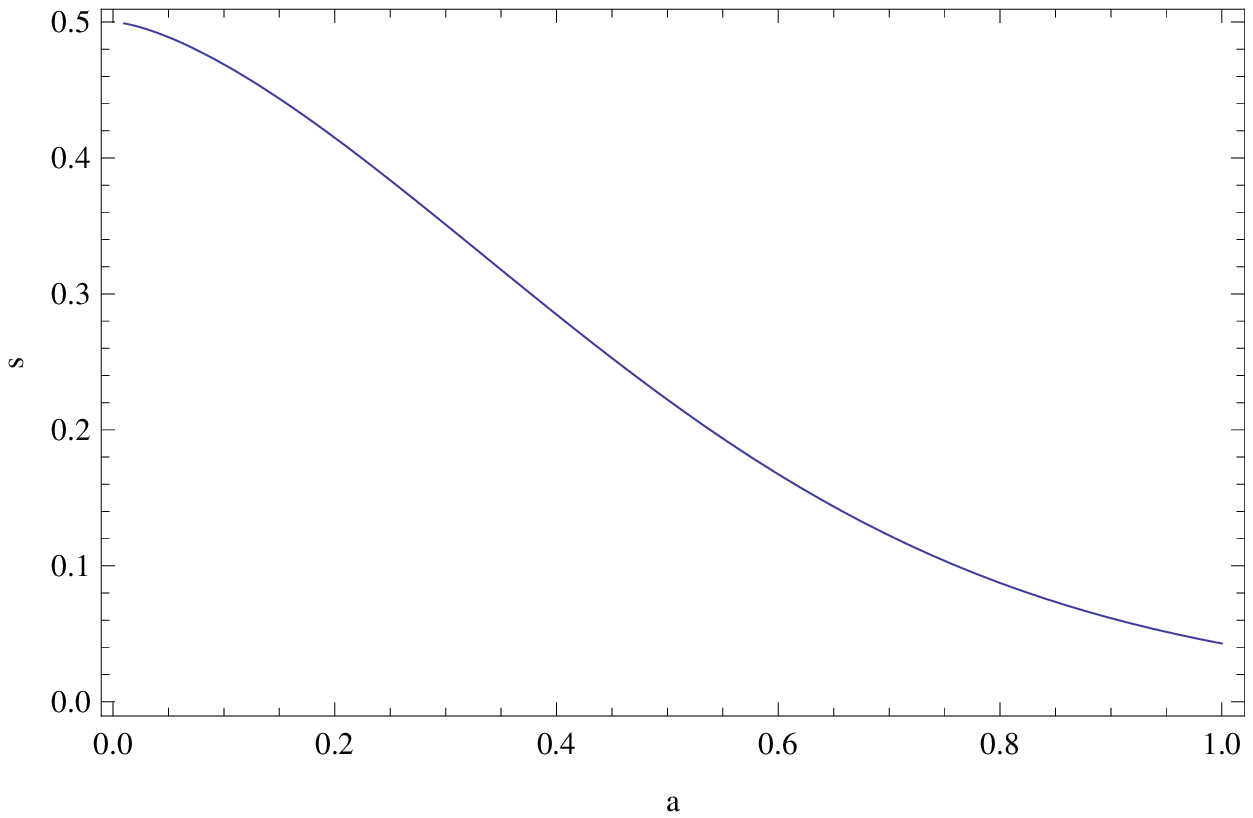}
\caption{Variation of statefinder pair parameters with $a$.}\label{fig9}
\end{figure}

\begin{figure}[thbp]
\centering
\includegraphics[width=0.35\textwidth]{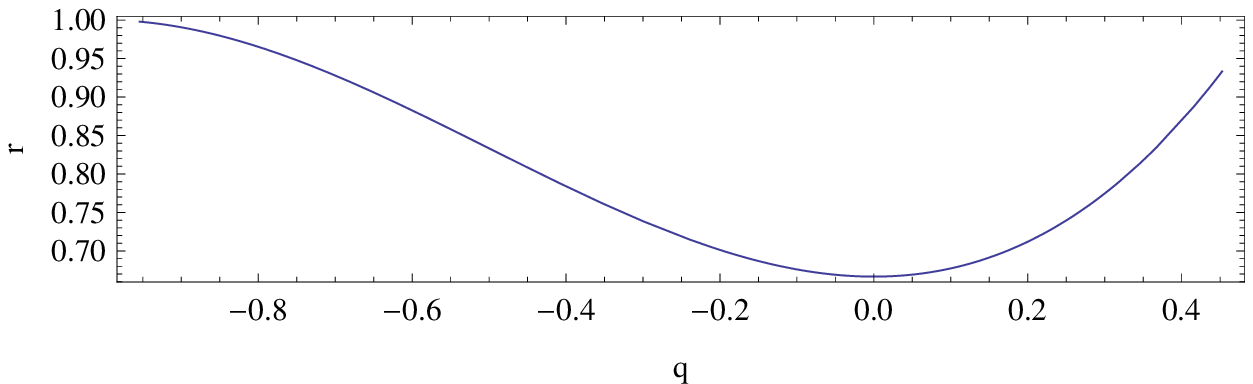}
\includegraphics[width=0.35\textwidth]{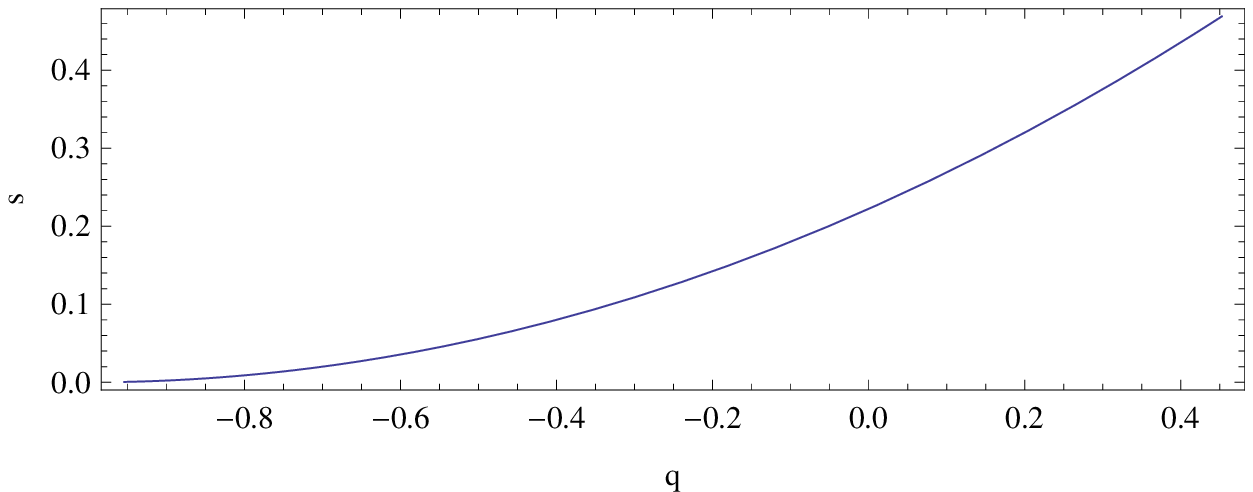}
\caption{Variation of statfinder pair parameters with deceleration parameter $a$.}\label{fig10}
\end{figure}

\begin{figure}[thbp]
\centering
\includegraphics[width=0.35\textwidth]{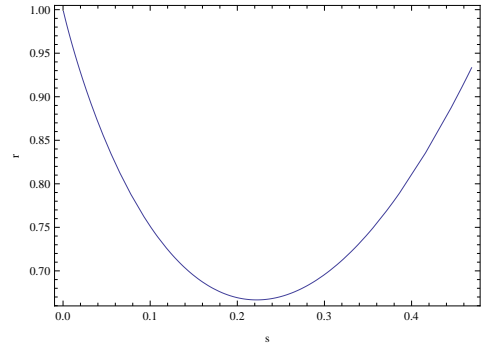}
\caption{Variation of statfinder parameters $r$ with $s$.} \label{fig11}
\end{figure}

The evolution of the statfinder pair parameters for GGDE in the framework of DGP braneworld have
been shown in Fig.~\ref{fig9}, Fig.~\ref{fig10} and Fig.~\ref{fig11}. From all of these figure we see that $r$ diverge, which corresponds
to the matter dominated Universe.

\section{Conclusion and Discussion}\label{sec7}

In the present study of generalized ghost dark energy model under
the framework of DGP braneworld scenario, we have tried to explore
several physical aspects of the model. From the observational
evidences it has been confirmed that almost three-fourth part of the
total energy of our universe is in the form of dark energy, which
is playing very crucial role to explain the current phase of our
universe. Among various available dark energy candidates, GGDE
model found out to be one of the most successful model as it is
free from the problem of any extra degree of freedom and it also
satisfies the stability criterion. In this section we are going to summarize
some of the interesting results that we have observed under this
present investigation.

(i) \textbf{Hubble Parameter}: Recent observational results
suggests that the value of Hubble parameter getting decreases with
the evolution of the universe. Here we have also studied the
variation of $H$ with respect to scale factor $a$ in Fig.~\ref{fig3} and
the variation shows a monotonically decreasing nature of $H$ with
$a$, which satisfies the observational results. From recent
observational data of Planck collaborators \cite{Planck2016}, the
value of Hubble constant can be obtained as  $h=67.8 \pm 0.9\
kms^{-1}\ Mpc^{-1} $. From Fig.~\ref{fig3} we also obtain the value of
Hubble constant  as $H\sim69$ for present time (i.e. $a=1$). Which
indicates that GGDE model completely agrees with the observational
value of Hubble constant.

(ii)\textbf{EoS Parameter}: The equation of state parameter
$\omega_D$ has been obtained for generalized ghost dark energy
model in DGP braneworld scenario. At the late time, the DE acts
like a cosmological constant due to asymptotic behavior of
$\omega_{D}$. From the Fig.~\ref{fig4}., we see that $\omega_{D}$ can
never cross -1, which is similar to quintessence behavior. EoS
parameter varies from zero at early time to $-1$ at late time.
Also the dimensionless GGDE density parameter $\Omega_{D}$ in
terms of scale factor $a$ is shown in Fig.~\ref{fig5}. We see that at the
early time, $\Omega_{D}\rightarrow 0$ and at late time,
$\Omega_{D}\rightarrow 1$, i.e. dark energy dominated.

(iii) \textbf{Deceleration Parameter}: At the early epoch of
evolution, the universe was dominated by matter, i.e. which caused
the decelerating phase of the universe. But later on due to
expansion of the universe, the phase flipped from deceleration to
acceleration. This flipping caused due to domination of dark
energy. Here we have studied the deceleration parameter and shows
its variation with respect to the scale factor in Fig.~\ref{fig6}, it has
been observed from this plot that the signature of the
deceleration parameter flipped at $a\sim 0.5$, which indicates the
transition of the universe from deceleration phase to acceleration
phase. This is a clear indication for physical applicability of
our present form of DE as GGDE.

(iv) \textbf{$\omega_{D}-\acute{\omega_{D}}$ Analysis}: We have
computed $\omega'_D$ in terms of $a$. We have
drawn $\omega'_D$ versus $\omega_D$ in Fig.~\ref{fig7}. In
$\omega_D$-$\omega'_D$ analysis, we have found only thawing region
in that plane because $(\acute{\omega_{D}}>0$ for $\omega_{D}<0)$.
So no freezing region available in our GGDE in DGP Model.

(v) \textbf{Adiabatic Sound Speed ($v_s)$}: The study of  adiabatic sound speed
is an important parameter to describe the stability. For any stable
solution we must have $0<v_s^2<1$. We have calculated the sound speed
for the model GGDE and showed the variation in Fig.~\ref{fig8}.
From these figure it has been found that $v_s^2$ remains positive and
within range for GGDE model.

(vi)\textbf{Statefinder Parameter}: Study of statefinder parameters is very essential
for any physically acceptable DE model. It plays an important role to discriminate among
various dark energy models. We have shown variations of the statefinder parameters in
Fig.~\ref{fig9}, Fig.~\ref{fig10} and Fig.~\ref{fig11}. From Fig.~\ref{fig11} the variation
of $r$ and $s$ in $r-s$ plane shows that the universe evolves and diverges from fixed
point, i.e. from SCDM ($s\approx 1,r\approx 1$) universe (steady state) and attained
least value then it increases and leads to $\Lambda CDM$ model with $(s=0,r=1)$. From this study one
can discriminate between the GGDE model with other DE  models.\\

As a final comment we can conclude that a set of physically acceptable
solutions for GGDE under the framework of DGP model has been obtained. Through the
analysis of various physical parameters we have found that our model is stable,
which confirms that generalized ghost dark energy is one of the most acceptable form
to describe accelerating phase of the present universe, whereas various studies on
GDE model were unable to provide a stable solution of the universe. Again in the
earlier work of Biswas et al.~\cite{biswas2018} on GGDE model have showed that
GGDE model provides a stable solution of the universe under the framework of FRW
universe, in the similar fashion we have found that our present study on GGDE model
also provides a set of stable solutions under DGP model of the universe.

\end{document}